\documentclass{vgtc}                          
\usepackage{xcolor}
\usepackage{soul}

\graphicspath{{figures/}{pictures/}{images/}{./}} 

\usepackage{times}                     

\usepackage{tabu}                      
\usepackage{booktabs}                  
\usepackage{lipsum}                    
\usepackage{mwe}                       
\usepackage{subcaption}
\usepackage{mathptmx}                  

\newcommand{\red}[1]
{{\leavevmode\color{black}#1}}

\onlineid{0}

\vgtccategory{Research}

\vgtcinsertpkg

\title{Scene Awareness While Using Multiple Navigation Aids in AR Search \thanks {This is the preprint version of the submitted 2-page summary document (non-archival) for a poster presented at IEEE ISMAR 2025. This work was initially made available on the author's personal website [yujnkm.com] in October 2025, and subsequently uploaded to arXiv for broader accessibility.}}

 \author{Radha Kumaran \textsuperscript{1}
 \and You-Jin Kim \textsuperscript{2}
 \and Emily Machniak \textsuperscript{1}
 \and Shane Dirksen \textsuperscript{1}
 \and Junhyung Yoon \textsuperscript{1}
 \and Tom Bullock \textsuperscript{1}
 \and Barry Giesbrecht \textsuperscript{1}
 \and Tobias Höllerer \textsuperscript{1}}

\affiliation{
\vspace*{-1.5ex} 
\textsuperscript{1} \scriptsize{University of California, Santa Barbara, USA}\\
\textsuperscript{2} \scriptsize{Texas A\&M University, USA}
\vspace*{-2ex} 
}

\abstract{
Augmented reality (AR) allows virtual information to be presented in the real world, providing support for numerous tasks including search and navigation. Allowing users access to multiple navigation aids may help leverage the benefits of different navigational guidance methods, but may also have negative perceptual and cognitive impacts. 
In this study, users performed searches for virtual gems within a large-scale augmented environment while choosing to deploy two different navigation aids either independently or simultaneously: world-locked arrows and an on-screen radar. After completing the search, participants were asked to recall objects that may or may not have been present in the scene.    
The use of navigation aids impacted object recall, with impaired recall of objects in the environment when an aid was switched on. The results point at possible impact factors of object awareness in mobile AR and 
underscore the potential for adaptable interfaces to support users navigating the physical world.
} 


\begin{document}


\firstsection{Introduction}

\maketitle

AR has the potential to be valuable for search and navigation support due to its ability to present virtual guidance (such as navigation aids) in the context of the real search environment. While participants may benefit from having access to multiple aids, it is important to consider potential costs in terms of increased perceptual and cognitive load. Increased visual clutter in AR could reduce users' awareness of their environment~\cite{zollmann_visualization_2021}, placing the user at risk if navigating unpredictable and hazardous environments, such as in an emergency search and rescue operation.

To investigate this issue, we designed 
a new study in a series of experiments exploring the influence of mobile AR interfaces on the recall of scene objects. Previous work reported that in wide area outdoor AR treasure hunts where participants searched for virtual gems \cite{kim_investigating_2022,kumaran_impact_2023}, participants' recall of physical scene objects was strongly impacted. In an indoor treasure hunt for virtual {\em and} physical gems, under varying scene augmentation density and controlled path guidance conditions, object recall did not significantly differ regarding physical vs. virtual objects, but participants experienced source confusion much more with physical objects than with virtual ones \cite{kim_go_2025}.

In this experiment, we aim to understand the impact of visual guidance tools on attention to the user’s environment, since there could be perceptual or cognitive impacts of AR information display. 
Here, participants searched for virtual gems using two navigation aids (2D on-screen radar and 3D in-world arrows), choosing either one (or none), or both in combination. In some experimental trials, the stability of the arrows was manipulated in order to understand the extent to which unreliability impacts effectiveness with, and preference for, world-locked navigation guidance. After each trial, participants’ memory of the environment was assessed through a recall test, where they indicated which objects from a provided list they remembered seeing. 



\begin{figure}[t]
    \centering
    \includegraphics[width=0.7\linewidth]{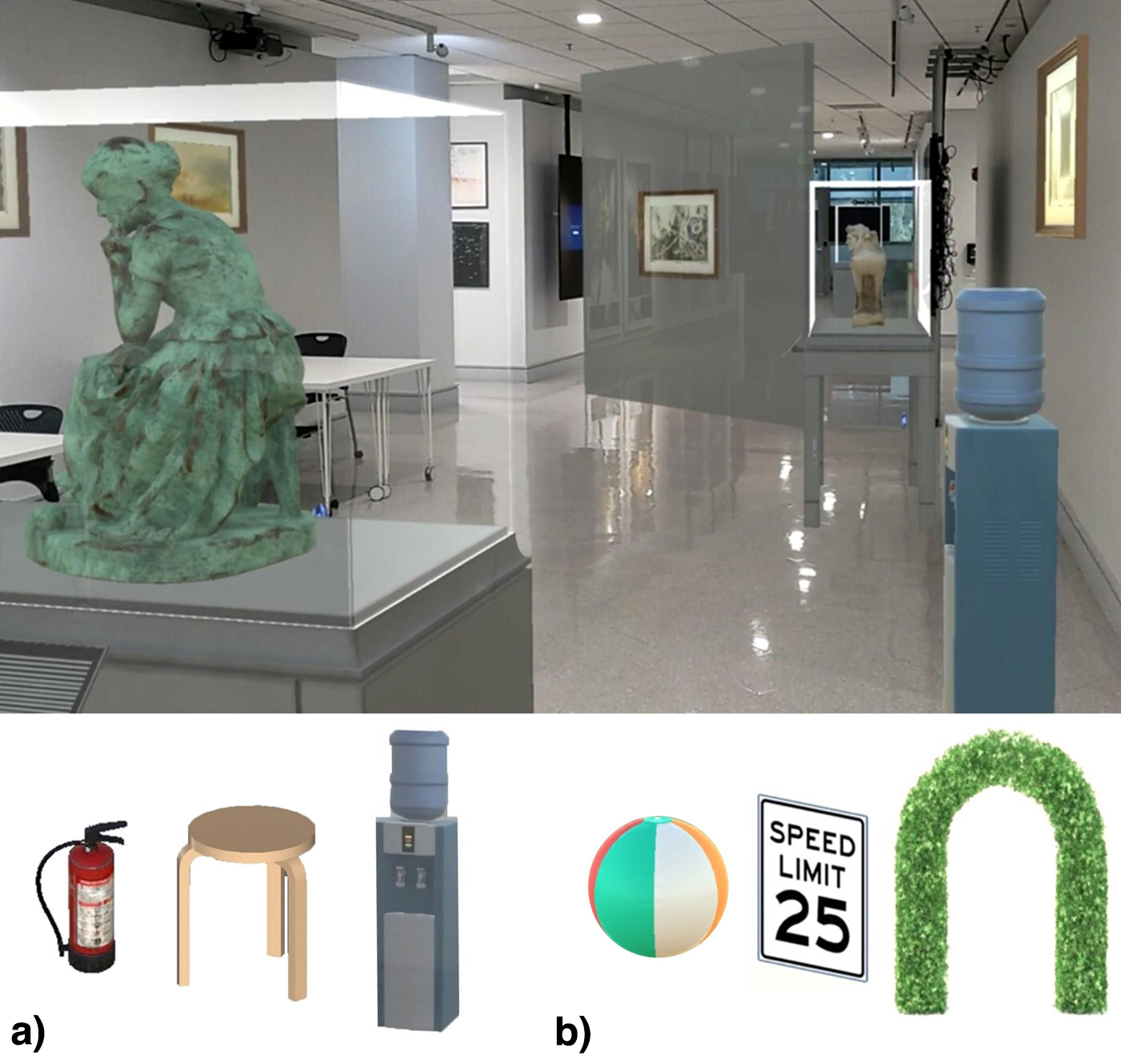} 
    \caption{Top: Experiment environment, Bottom: Some small, medium, and large (left to right) objects included in the recall tests. a) semantically congruent objects. b) semantically incongruent objects. Overall, our set of candidate objects for recall questions consisted of 18 virtual object models and another 18 names of objects that were always absent from the scene.}
    \label{fig:recall-objects}
    \vspace*{-3ex}
\end{figure}

\section{Experiment}

The 24 participants completed 9 experimental trials, and were tested on their recall of a unique set of objects after each trial. In each trial, participants performed a primary gem search task, and were instructed to find as many of 12 distributed gems as they could in the museum-themed indoor search space (approximately 208 sq.m). The presence of a \textit{navigation aid} that guided participants to the locations of the gems was manipulated within-subjects, implemented as 3D in-world arrows and/or a 2D on-screen radar in our experiment. In some trials, the \textit{stability} of the 3D arrows was also manipulated with simulated display latency, implemented using a low-pass IIR filter and a complete displacement of arrows in some sections of the trials.

The first 3 \textit{baseline} trials were performed with either no navigation aid or a single aid that was always on (arrows or radar). In the next 6 \textit{mixed} trials, participants were allowed to control the visibility of each aid independently of the other, and could also use both simultaneously if desired. In the mixed trials, the stability of the arrows was manipulated at 3 levels (none, mild or severe instability), which were chosen based on pilot experiments.

After each trial, participants were given a list of four objects, and indicated which of them were present in the scene. Of the four objects, two had been actually present, and two had not. Each set of objects was only present in a single trial. The order of the objects appearing in the 9 trials was randomized using Latin Square counterbalancing. In each of these pairs of objects, one was semantically congruent with the scene and the space, and one was not. To decide congruence, three experimenters rated the objects on a scale of 0-10 for semantic and aesthetic congruence with the space and the environment, and the scores were averaged. Objects that scored less than 3/10 were considered incongruent, and objects that scored more than 7/10 were considered congruent.

After the experiment, participants also participated in a semi-structured interview regarding their experiences.

All statistical analysis was performed in R and variables of interest were modeled as generalized linear mixed models (GLMMs) with a binomial distribution. GLMMs were implemented using the {\small lme4} package, or the {\small glmmTMB} package in case of overdispersed binomial variables (in this paper, the trial order analysis). Post-hoc pairwise comparisons were performed using estimated marginal means with Bonferroni adjustments from the {\small emmeans} package.

\section{Results}

\begin{figure}[t]
    \centering
    \begin{subfigure}{0.48\columnwidth}
        \includegraphics[width=\linewidth]{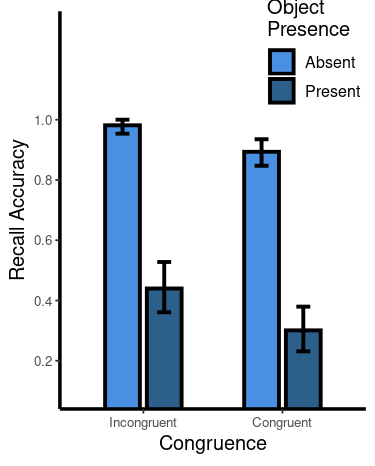} 
        \caption{}
        \label{fig:recallgt}
    \end{subfigure}\hfill
    \begin{subfigure}{0.48\columnwidth}
        \includegraphics[width=\linewidth]{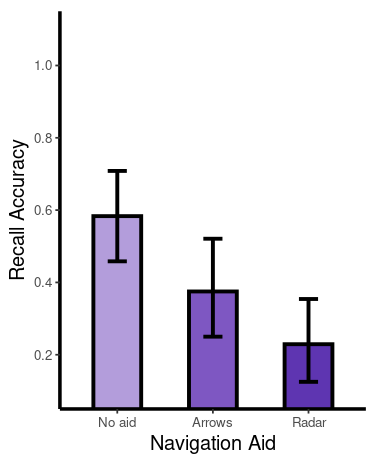} 
        \caption{}
        \label{fig:recallmode}
    \end{subfigure}\hfill
    \caption{(a) Object recall accuracy as a function of semantic congruence and object presence. Participants were worse at remembering objects that were present compared to identifying absent objects.  Error bars are 95\% CI.
    (b) recall accuracy (for objects present in the scene only) as a function of the three baseline conditions. Recall was worse with the radar compared to no aid. }
    \label{fig:recall}
    \vspace*{-3ex}
\end{figure}

A binomial GLMM with object presence (present/absent), object congruence (congruent/incongruent) and trial order (1-9) as fixed effects was fit to assess object recall (see Figure \ref{fig:recallgt}), with participant ID included as a random effect. There were main effects of object congruence ($\chi_{}^{\mathrm{2}}$(1) = 4.72, $p$ = 0.029) and object presence ($\chi_{}^{\mathrm{2}}$(1) = 13.19, $p$ = 0.0003), but no effect of trial order or interactions (all $\chi^2$ $<$ 0.86, all $p$ $>$ 0.05). Incongruent objects were recalled more accurately than congruent objects [$\beta$  = 1.24, $SE$ = 0.30, $p$ $<$ 0.0001], and recall accuracy was lower for objects that were actually present in the environment compared to absent objects [$\beta$  = -3.82, $SE$ = 0.31, $p$ $<$ 0.0001]. The theoretical marginal $R^2$ of the model was 0.872, and the conditional $R^2$ was 0.952. \red{This suggests that participants likely presumed objects to be absent unless they were sure they had seen it.}

For the objects that were actually present in the scene, a binomial GLMM with navigation aid and object congruence as fixed effects (and participant ID as a random effect) indicated an impact of navigation aid among the three baseline conditions ($\chi_{}^{\mathrm{2}}$(2) = 6.949, $p$ = 0.031), with significantly worse recall for the radar than for no aid [$\beta$  = -1.87, $SE$ = 0.54, $p$ = 0.002] as indicated in Figure \ref{fig:recallmode}. None of the other pairwise comparisons were significant [all $\beta$ $<$ 1.86, all $p$ $>$ 0.05], and there was no main effect of object congruence or interaction (all $\chi_{}^{\mathrm{2}}$ $<$ 1.43, all $p$ $>$ 0.05). The theoretical marginal $R^2$ of the model was 0.249, and the conditional $R^2$ was 0.27.

Surprisingly only 14 out of the 24 participants reported switching on and off the navigation aids during the task, with 8 choosing to keep both on for the entirety of the task even if the aids were only occasionally helpful (``... the arrows were occasionally a bit distracting... occasional moments [when] [the arrows] benefited me, then occasional moments where it wasn’t super helpful'', P8) or not useful at all (``I kept both of them on... even when both were turned on, I didn’t really use the arrows'', P26). Participants also mentioned finding it harder to pay attention to the environment when using the navigation aids (``... once you have the radar you simply follow the radar. You don’t look at the objects.'', P26), which would explain the impaired recall of objects with the radar conditions even though participants were primed to the object recall task.

\section{Discussion}

Many participants kept all available AR navigation guidance visible, even when the arrows were not helpful and at times, distracted from the task. The results of the object recall test and participant interviews indicate that the use of navigation aids limited their ability to attend to their environment (even though explicitly instructed to). This indicates a general tendency to over-reliance on all available navigation aids, even when the arrows were clearly perceived to be
unstable and inaccurate,  which is particularly concerning for wide-area AR tasks that require environment scanning for navigation, since we observed reduced attention to the environment with radar use compared to no navigation aid.

Although similar effects have been previously reported for general tool use (e.g. \cite{stanovich2018miserliness}), it is surprising in the context of this experiment because of the amount of instability-induced visual distraction participants were willing to tolerate in a task that required them to actively scan and navigate a large environment. 

It is therefore important to control the amount of visual information a user is exposed to when attention to the environment is important, and context-adaptive cueing systems could help address this by focusing user attention in busy or unsafe environments, thus allowing users to gain awareness of their surroundings.

\section*{Acknowledgments}
This work was supported in part by  
contract W911NF-19-2-0026 from the U.S. Army Research Office, ONR grant
N00014-23-1-2118, as well as NSF grant IIS-2211784.


\bibliographystyle{abbrv-doi-narrow}

\bibliography{template}
\end{document}